\documentclass[a4paper,fleqn]{cas-sc}
\usepackage[numbers]{natbib}
\usepackage{subfigure}
\usepackage[ruled]{algorithm2e}
\usepackage{bm}

\usepackage{caption}
\usepackage{amssymb}
\usepackage{graphicx}
\usepackage{float} 

\usepackage{makecell}
\usepackage{multirow}

\usepackage{cleveref}
\usepackage{color}

\usepackage{hyperref}
\hypersetup{hypertex=true,
            colorlinks=true,
            linkcolor=blue,
            anchorcolor=blue,
            citecolor=blue}

\usepackage{amsthm}
\newtheorem{definition}{Definition}%
\newtheorem{theorem}{Theorem}

\def\tsc#1{\csdef{#1}{\textsc{\lowercase{#1}}\xspace}}
\tsc{WGM}
\tsc{QE}
\tsc{EP}
\tsc{PMS}
\tsc{BEC}
\tsc{DE}
\usepackage{caption}
\begin{document}
\let\WriteBookmarks\relax
\def\floatpagepagefraction{1}
\def\textpagefraction{.001}
\let\printorcid\relax
\shortauthors{Linlin Wang et~al.}

\title [mode = title]{Fiber Signal Denoising Algorithm using Hybrid Deep Learning Networks}                  
\tnotemark[1]

\author[1]{Linlin Wang}[style=chinese]
\credit{Conceptualization, Methodology, Formal analysis, Visualization, Writing - original draft}
\address[1]{School of Mathematics, Renmin University of China, Beijing 100872, China}

\author[1]{Wei Wang}[style=chinese]
\credit{Methodology, Validation, Writing - review \& editing}

\author[3]{Dezhao Wang}[style=chinese]
\credit{Conceptualization, Data curation}
\address[3]{Beijing Jhbf Technology Development Co., Ltd., China}

\author[1]{Shanwen Wang}[style=chinese]
\cormark[1]
\credit{Conceptualization, Project administration, Funding acquisition, Supervision, Writing - review \& editing}
\ead{s_wang@ruc.edu.cn}

\tnotetext[1]{Linlin Wang, Wei Wang, and Shanwen Wang were funded by Tianjin Yunhong Technology Development Grant No. 2021020531.}

\cortext[1]{Corresponding author.}

\begin{abstract}
    With the applicability of optical fiber-based distributed acoustic sensing (DAS) systems, effective signal processing and analysis approaches are needed to promote its popularization in the field of intelligent transportation systems (ITS). This paper presents a signal denoising algorithm using a hybrid deep-learning network (HDLNet). Without annotated data and time-consuming labeling, this self-supervised network runs in parallel, combining an autoencoder for denoising (DAE) and a long short-term memory (LSTM) for sequential processing. Additionally, a line-by-line matching algorithm for vehicle detection and tracking is introduced, thus realizing the complete processing of fiber signal denoising and feature extraction. Experiments were carried out on a self-established real highway tunnel dataset, showing that our proposed hybrid network yields more satisfactory denoising performance than Spatial-domain DAE.
\end{abstract}


\maketitle

\section{Introduction}

Emerging optical fiber-based distributed acoustic sensing (DAS) systems utilize advanced $\Phi$-OTDR technology to detect temperature and strain changes along fiber locations \cite{taylor1993apparatus}. This system can realize real-time and weather independent traffic state perception and have a broad application prospect, owing to its high sensitivity, cost-effectiveness, durability and many other design advantages \cite{wang2019comprehensive}. Since the feasibility of using fiber vibrations to extract road traffic information was first demonstrated in \cite{martin2016interferometry}, many scholars have begun to explore the applications of DAS systems in transportation, such as vehicle detection \cite{wellbrock2019first}, \cite{kowarik2020fiber}, traffic flow detection \cite{liu2018traffic}, \cite{wiesmeyr2021distributed}, vehicle classification \cite{liu2019vehicle}, etc. 

Limited by the means of signal acquisition, transmission and storage, there are massive noise, outliers and other problems in fiber signals, causing trouble to reveal vibration information \cite{dou2017distributed}. This has become the bottleneck of signal analysis and modeling. Traditional denoising methods like wavelet denoising \cite{liu2018traffic}, \cite{wu2015separation}, \cite{huot2017automatic} hit a bottleneck in terms of processing responses in the age of big data. In recent years, deep learning algorithms have made great progress in many areas \cite{zhan2022evolutionary}, offering advantages in processing speed as well as flexibility. As an important subset, neural networks have shown high efficiency and great potential in many specific tasks \cite{ilesanmi2021methods}. Some deep learning algorithms are applied in fiber signal denoising \cite{wang2021rapid}, \cite{van2022deep}, \cite{yuan2023spatial}. However, none of these methods pay attention to highly spatial and temporal correlation of optical fiber signals. Besides, after data preprocessing, it is of great importance to design object detection and multiple-object tracking algorithms for vehicle positioning and tracking. Unfortunately, many scholars only viewed vehicle trajectories as straight lines and utilized linear detection methods, such as Hough transform \cite{wiesmeyr2021distributed} and Radon transform \cite{wang2021vehicle}. More accurate algorithms need to be designed for dealing with cluttered or crossed trajectories.

\begin{figure}[!t]
    \centering
    \includegraphics[width=3.6in]{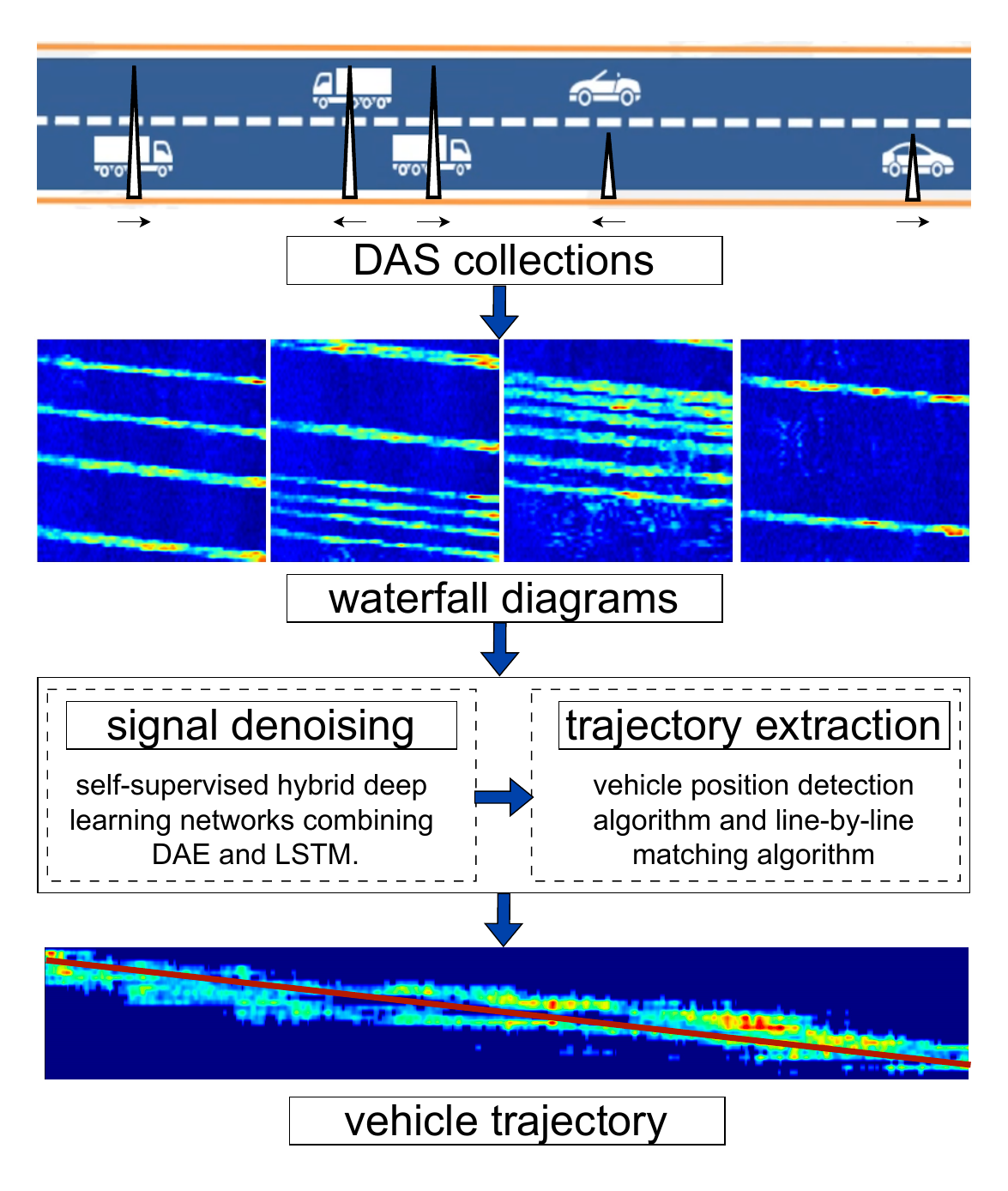}
    \hfill
    \caption{An illustration of our framework for processing optical fiber signals recorded by DAS systems.}
    \label{sum}
\end{figure}

This paper is devoted to promote the advantages of DAS systems by studying the key technologies of signal denoising. A complete set of algorithm flow for fiber signal processing is presented and the overall framework is illustrate in Fig. \ref{sum}. In the phase of signal denoising, a novel self-supervised hybrid deep-learning network (HDLNet) combining a denoising autoencoder (DAE) \cite{lee2021noise} and a long short-term memory (LSTM) \cite {hochreiter1997long} is proposed. As one variant of autoencoders, DAE has an ability to reconstruct original noise-free data from noisy inputs. Considering real scenario settings, the convolution kernel is described by physical Flamant-Boussinesq proximation \cite{Fung1966FoundationOS}. The architecture of our DAE is the lightweight but effective U-Net model, which consists of a symmetrical encoder and decoder structure. Particularly, it has a skip structure to pass features from the encoder path to the decoder path \cite{ronneberger2015u}. Besides, considering the high spatial and temporal correlation of optical fiber signals, LSTM is supplemented to process time series signals. It controls the flow of information by introducing three gating mechanisms (input gate, forget gate, output gate), enabling the network to remember information for a long time. LSTM exhibits powerful sequence processing capabilities in many fields such as natural language processing \cite{yin2017comparative} and speech recognition \cite{graves2013hybrid}. 

% In the phase of feature extraction, optical fiber signals are represented as three-dimensional waterfall diagrams. Then we propose a vehicle position detection algorithm and a vehicle trajectory extraction algorithm to achieve detection and overall tracking for vehicles. 

This research is of great significance for developing new applications of distributed optical fiber sensing technology in the field of intelligent highway traffic detection. Without subjective and error-prone manual annotations, the proposed self-supervised framework can produce high-quality results for denoising tasks. Besides, the performance of the proposed algorithms is verified using real self-established optical fiber datasets from highway tunnel scenarios. After denoising, we utilize vehicle position detection and trajectory extraction algorithms to achieve overall tracking for vehicles \cite{WANG2025104228}. Experimental results confirm that our denoising algorithm is more effective than Time-domain DAE \cite{van2022deep} and Spatial-domain DAE \cite{yuan2023spatial}. 
% And the vehicle trajectory extraction algorithm proposed in this paper can achieve multi-target separation and whole-process tracking. 
In conclusion, our proposed denoising HDLNet combining DAE and LSTM can achieve ideal denoising effects under the premise of considering spatio-temporal correlations. And the overall algorithms shows our effectiveness in practical highway scenarios. 

% The main contributions of this paper are as follows:
% \begin{enumerate}
%     % \item To the best of our knowledge, we are the first to provide datasets using optical fiber-based DAS systems in highway tunnel scenarios.
%     \item The proposed denoising HDLNet combining DAE and LSTM can achieve ideal denoising effects under the premise of considering spatio-temporal correlations.
%     \item A whole optical fiber processing algorithm is presented, including signal denoising and subsequent trajectory extraction. This can be applied and complemented in intelligent transportation systems.
% \end{enumerate}

The remainder of this paper is as follows. Section \ref{related_works} recalls signal denoising algorithms for DAS technology. Section \ref{method} displays the methodology and physical model. Section \ref{architecture} introduces the design of denoising HDLNet architecture. Section \ref{experiments} exhibits our self-established datasets, implementation details and experimental results. Section \ref{conclusion} provides conclusions and limitations of this study.

\section{Related Works}
\label{related_works}

There are mainly two approaches for denoising optical fiber signals collected by DAS systems. One is to analyze one-dimensional signals, either time or spatial series signals; the other is to analyze two-dimensional matrix by accumulating spatial series signals along the time axis.

\subsection{One-dimensional signal denoising}

Traditional denoising methods include time-domain filtering and frequency-domain filtering. Time-domain filtering simply separates signal and noise by a sliding window directly on time series\cite{lee2002general}, such as moving average filtering \cite{alvarez2005detrending}, median filtering \cite{justusson2006median}, adaptive filtering \cite{diniz1997adaptive}, etc. Although variations of these time-domain filtering methods are applicable to non-linear non-stationary data, they cannot protect the locally detail changes of signals. In contrast, frequency-domain filtering decomposes signals by transforming them from the time domain to the frequency domain \cite{shynk1992frequency}, such as Wiener filter \cite{gardner1993cyclic}, Kalman filter \cite{welch1995introduction}, low-pass filter \cite{kaiser1977data}, among others. However, the limitations of frequency-domain filtering in the processing of non-smooth or abrupt signals are gradually becoming apparent.

Due to the advantages of wavelets in both time and frequency domain, many scholars have begun to pay attention to its application in the field of denoising. A wavelet threshold denoising algorithm combining soft and hard thresholds was proposed in \cite{liu2018traffic}. Continuous wavelet transform and clustering algorithm were applied to identify different types of noise \cite{huot2017automatic}. A six-level wavelet multi-scale decomposition algorithm based on discrete-time wavelet transform was adopted \cite{wu2015separation}. However, these denoising methods have certain limitations because the lack of rapid response capabilities.

\subsection{Two-dimensional signal denoising}

Two-dimensional signal denoising focuses on image denoising of waterfall diagrams. This problem was viewed as an image defuzzification problem in \cite{van2022deep}, with the fast iterative shrinkage thresholding algorithm utilized to improve image resolution. Although noise can be removed to a certain extent, this method still lacks real-time processing. Deep learning algorithms have made remarkable achievements in image denoising these years \cite{tian2020deep}. Neural networks can achieve higher denoising accuracy by using a large amount of data for training. Artificial neural networks were first introduced into the field of optical fiber signal denoising in \cite{kowarik2020fiber} and achieved clearly visible outcomes. An attention-based convolutional neural network was proposed in \cite{wang2021rapid} to approximate classical methods (wavelet transform and S-transform) results. DAE was first utilized in \cite{van2022deep} and further modified from time-DAE to spatial-DAE in \cite{yuan2023spatial}. But currently deep-learning based algorithms do not take into account the serial correlation of spatio-temporal signals, ignoring the intrinsic relationship. 

\section{Method}
\label{method}

\subsection{Problem Formulation}

Denote the original noiseless image as $x = (x_1,...,x_{N_t}) \in M_{N_d \times N_t}(\mathbb{R})$ and observed noisy image with noises $\epsilon \in M_{N_d \times N_t}(\mathbb{R})$ as $y = (y_1,...y_{N_t})\in M_{N_d \times N_t}(\mathbb{R})$. According to \cite[Chapter 5]{rafael2017digital}, the observed image $y$ can be composed of $y = f(x) + \epsilon$, where $f(\cdot)$ is the degenerate function. In this article, we assume the function $f(\cdot)$ is linear:
\begin{equation}
    y = Ax + \epsilon,
\end{equation}
where $A \in M_{N_d \times N_d}(\mathbb{R})$ is the degenerate/observation matrix, and has the form of $(k,..., k)^{T}, k \in \mathbb{R}^{N_d}$. The goal of image denoising is to reconstruct the restored image $\hat{x} \in M_{N_d \times N_t}(\mathbb{R})$. Therefore, the image denoising problem can be transformed into the least absolute shrinkage and selection operator (LASSO) problem \cite{ranstam2018lasso}:
\begin{equation}
    \hat{x_i} = \arg\min\limits_{x \in \mathbb{R}^{n}} \Vert Ax_i - y_i \Vert _2^2 + \lambda \Vert x_i \Vert _1,
\label{lasso}
\end{equation}
where $\lambda \geq 0$ is a regularization parameter.

In the field of image denoising, higher frequency component corresponds to the noise or detail part of the image, while the lower frequency component corresponds to the main structure of the image \cite{goyal2020image}. The Fourier transform can be applied here to convert signals from time domain to frequency domain, which can reveal the frequency component of signals \cite{allen2004signal}. Below we give the Fourier transform formula to explain, and the signal collected in actual reality needs to be discretized.

% \begin{assumption}
%     $f(t)$ satisfies the Dirichlet condition on any finite interval. 
%     \begin{enumerate}
%         \item $f(t)$ is continuous or has only a finite number of discontinuity point of the first kind.
%         \item $f(t)$ has at most a finite number of extreme points.
%     \end{enumerate}
% \end{assumption}

% \begin{assumption}
%     $f(t)$ is absolutely integrable on the infinite interval $(-\infty, +\infty)$
% \end{assumption}

\textbf{Notation:} In this subsection, we use lowercase variables (e.g., $x$ and $k$) to denote signals in the time domain, and corresponding uppercase counterparts (e.g., $X$ and $K$) to denote their spectra in the frequency domain.

\begin{definition}[Discrete-time Fourier transform]
    Denote $X$ and $K$ the discrete-time fourier transform (DTFT) of $x$ and $k$ respectively:
    \begin{equation}
        X(\omega) = \sum_{j = 1}^{n}x_je^{-i\omega j},
    \end{equation}
    \begin{equation}
        K(\omega) = \sum_{j = 1}^{n}k_je^{-i\omega j},
    \end{equation}
    where $i = \sqrt{-1}$ is imaginary unit and $\omega \in [-\pi, \pi]$ is angular frequency.
\end{definition}

% \begin{theorem}[Fourier integral theorem]
%     If $f(t)$ satisfies the above assumptions, then at the continuous points, we have
%     \begin{equation}
%         f(t) = \frac{1}{2\pi}\int_{-\infty}^{\infty}\left\{\int_{-\infty}^{\infty}f(\tau)e^{-i\omega \tau}d\tau\right\}e^{i\omega t}d\omega,
%     \end{equation}

% \end{theorem}

% \begin{definition}[Fourier transform]
%     Denote $X(omega)$ and  is the Fourier transform of $f(t)$:
%     \begin{equation}
%         F(\omega) = \int_{-\infty}^{\infty}f(t)e^{-i\omega t}dt.
%     \end{equation}
% \end{definition}

% \begin{definition}[Inverse Fourier transform]
%     $f(t)$ is the inverse Fourier transform of $F(\omega)$:
%     \begin{equation}
%         f(t) = \frac{1}{2\pi}\int_{-\infty}^{\infty}F(\omega)e^{i\omega t}d\omega.
%     \end{equation}
% \end{definition}

% For simplicity, we write this pair of Fourier transforms as $f(t) \leftrightarrow F(\omega)$. 

% According to the Frequency-Domain Convolution Theorem shown below: 
The frequency-domain convolution theorem states that the convolution of the observed noisy image $x$ and convolution kernel $k$ in the time domain is equivalent to their multiplication in the frequency domain. 
% \begin{theorem}[Time-Domain Convolution Theorem]
% If $f_1(t) \leftrightarrow F_1(\omega)$ and $f_2(t) \leftrightarrow F_2(\omega)$, then we have
%     \begin{equation}
%         f_1(t) * f_2(t) \leftrightarrow F_1(\omega)F_2(\omega).
%     \end{equation}
% \end{theorem}

% \begin{theorem}[Frequency-Domain Convolution Theorem]
%     If $f_1(t) \leftrightarrow F_1(\omega)$ and $f_2(t) \leftrightarrow F_2(\omega)$, then we have
%     \begin{equation}
%         f_1(t)f_2(t) \leftrightarrow \frac{1}{2 \pi}F_1(\omega) * F_2(\omega)
%     \end{equation}
% \end{theorem}

\begin{theorem}[Frequency-Domain Convolution Theorem]
    Denote $\mathcal{F}[x \cdot k]$ is the convolution of $x$ and $k$ in the time domain, then
    \begin{equation}
        \mathcal{F}[x \cdot k](\omega) = \frac{1}{2 \pi}X(\omega) *  K(\omega).
    \end{equation}
\end{theorem}

If we want to restore the observed noisy image based on a preset convolution kernel $k$, we only need to perform the Fourier transform and multiply them. The equivalent transformation of the LASSO problem in Eq. (\ref{lasso}) from the frequency domain is:
\begin{equation}
    \hat{X} = \arg\min\limits_{X \in \mathbb{R}^{n}} \Vert X * K - Y \Vert _2^2 + \lambda \Vert X \Vert _1.
\label{lasso_fre}
\end{equation}

It's worth noting that the convolution direction between $X$ and $K$ is along the sensor axis instead of time axis to achieve better denoising performance, as confirmed in \cite{yuan2023spatial}.

\subsection{Simulated DAS response}
\label{Simulated DAS response}

In this section, the physical Flamant-Boussinesq approximation model \cite{Fung1966FoundationOS} was utilized to derive the convolution kernel.

\subsubsection{Flamant-Boussinesq approximation}

Denote a given position in three-dimensional space as $(d_x, d_y, d_z)$, where $d_x$ and $d_y$ are the tangential and perpendicular distances to the road respectively, and $d_z$ is the laying depth beneath the ground. The quasi-static or geodetic deformation $P(d_x,d_y,d_z)$ at this position is:
\begin{equation}
    P(d_x,d_y,d_z) = \frac{F}{4 \pi G} \times p(d_x,d_y,d_z),
\end{equation}
where $F$ is the total force, $G$ is the uniform shear modulus, and $p(d_x,d_y,d_z) = \frac{d_x}{r^2}(\frac{d_z}{r} + \frac{2 \nu - 1}{1 + d_z/r})$ with $r = \sqrt{d_x^2+d_y^2+d_z^2}$ and Poisson's ratio $\nu$. 

% \begin{equation}
%     P(x,y,z) = \frac{F}{4 \pi G} \times \frac{x}{r^2}(\frac{z}{r} + \frac{2 \nu - 1}{1 + \frac{z}{r}})
% \end{equation}

Consequently, the expression for the equivalent DAS measurement can be found as:
\begin{equation}
k = \frac{1}{l}|P(d_x-\frac{l}{2}, d_y, d_z) - P(d_x+\frac{l}{2}, d_y, d_z)|,
\end{equation}
where $l$ is the DAS gauge length. After applying DTFT from $k$ to $K$, it can be interpreted as the impulse response of optical fibers, serving as the kernel function in Eq. (\ref{lasso_fre}).

% \begin{equation}
% \begin{aligned}
%     \frac{\partial k}{\partial y} &= \frac{1}{d}\left[ 3yz (\frac{x - \frac{l}{2}}{r_2^5} - \frac{x + \frac{l}{2}}{r_1^5})\right] \\
%     &+ (2 \nu - 1)y\frac{(x - \frac{l}{2})(2r_2+z)}{r_2(r_2^2+2r_2)^2}\\
%     &- (2 \nu - 1)y\frac{(x + \frac{l}{2})(2r_1+z)}{r_1(r_1^2+2r_1)^2}
% \end{aligned}
% \end{equation}

% \begin{equation}
%     \begin{aligned}
%         \frac{\partial k}{\partial F} &= \frac{1}{4 \pi Gd}\left[\frac{x + \frac{l}{2}}{r_2^2}(\frac{z}{r_2} + \frac{2 \nu - 1}{1 + \frac{z}{r_2}})\right]\\
%         &- \frac{1}{4 \pi Gd}\left[\frac{x + \frac{l}{2}}{r_1^2}(\frac{z}{r_1} + \frac{2 \nu - 1}{1 + \frac{z}{r_1}})\right]
% \end{aligned}
% \end{equation}
% where 
% \\$r_1 = \sqrt{(x + \frac{l}{2})^2+y^2+z^2}$ 
% \\and $r_2 = \sqrt{(x - \frac{l}{2})^2+y^2+z^2}$

\subsubsection{Practical Calculation}

In the actual calculation process, each vehicle has four wheel positions that generate geodetic deformation. Assuming the vehicle's axle length and wheelbase as $a$ and $b$ respectively, we can then calculate the geodetic deformation at each of the four wheels. Impulse response $k$ can be defined as the difference between the response $k_{x_1}$ at the front end of the vehicle $d_{x_1} = d_x + \frac{l}{2}$ and the response $k_{x_2}$ at the rear end of the vehicle $d_{x_2} = d_x - \frac{l}{2}$:
\begin{equation}
   k = |k_{x_2} - k_{x_1}|,
\label{impulse}
\end{equation}
where
\begin{equation}
    k_{x_1} = \sum_{i=1}^{4} w_i \cdot P(d_{x_1} + \alpha_i, d_y + \beta_i, d_z),
    \end{equation}
\begin{equation}
    k_{x_2} = \sum_{i=1}^{4} w_i \cdot P(d_{x_2} + \alpha_i, d_y + \beta_i, d_z).
\end{equation}

Here, $w_i$ represents the weight of the $i$-th wheel, and $(\alpha_i, \beta_i)$ represents the coordinates of the $i$-th wheel relative to the vehicle center. From left front to right rear of a vehicle, $(\alpha_i, \beta_i)$ are $(\frac{b}{2}, \frac{a}{2}), (\frac{b}{2}, -\frac{a}{2}), (-\frac{b}{2}, -\frac{a}{2}) (-\frac{b}{2}, \frac{a}{2})$.

\begin{figure*}[htbp]
    \centering
    \subfigure[$d_y$ change]{
        \includegraphics[width=2.5in]{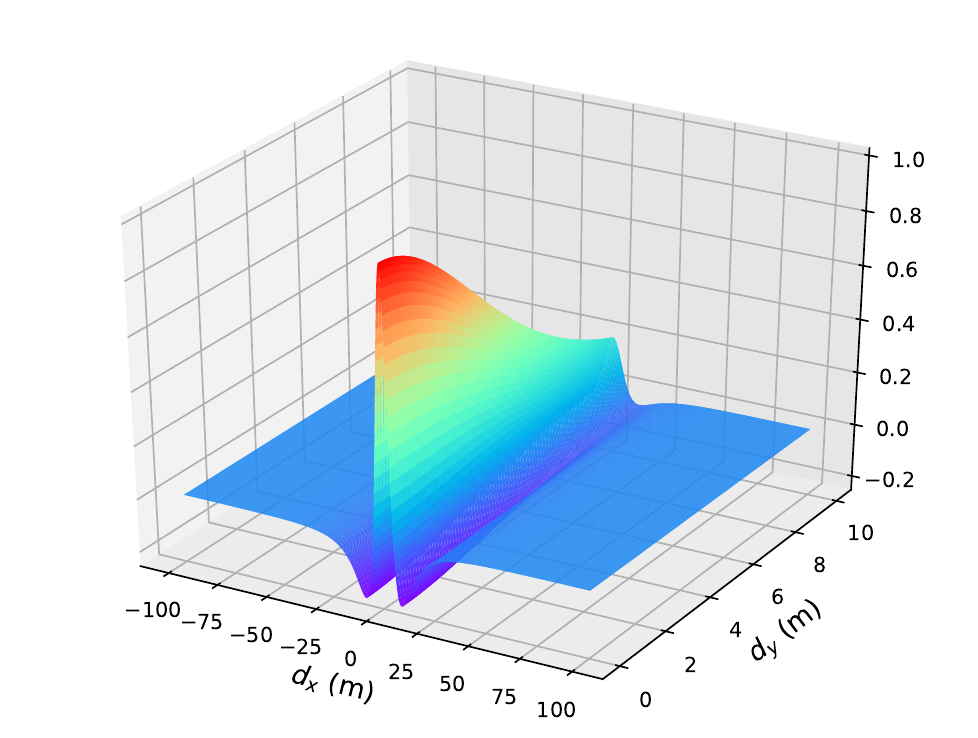}}
    \subfigure[$F$ change]{
        \includegraphics[width=2.5in]{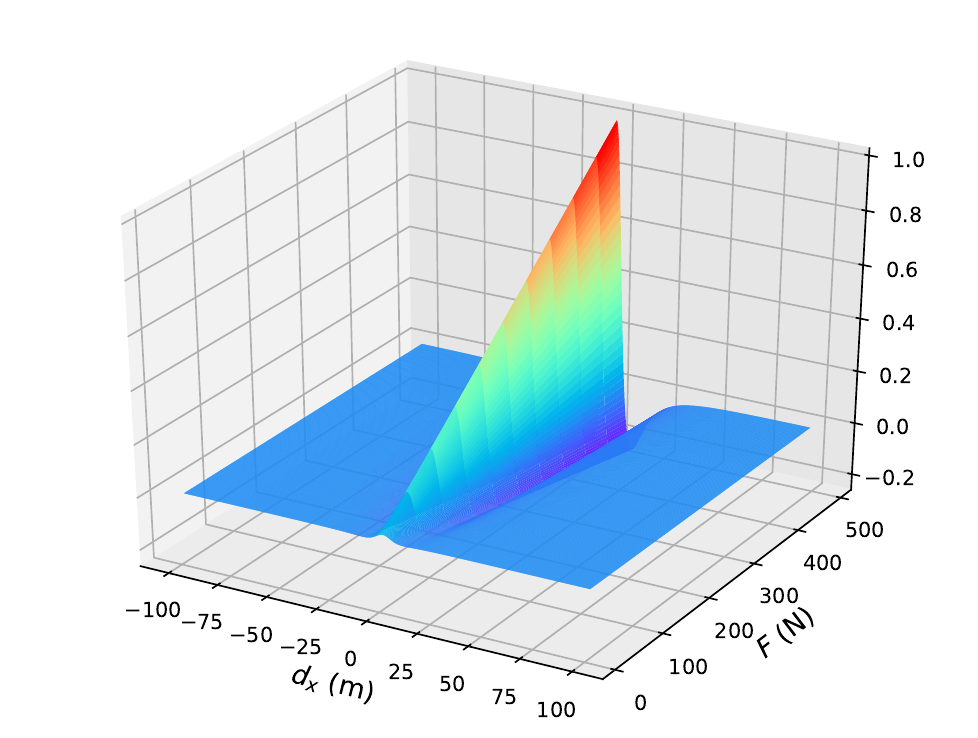}}
    \hfill
    \caption{Numerical simulation of quasi-static impulse responses. (a) Various road-perpendicular offsets $d_y$ to the fiber, (b) various vehicle loads $F$.}
    \label{F-B}
\end{figure*}

\begin{figure*}[!t]
    \centering
    \includegraphics[width=5in]{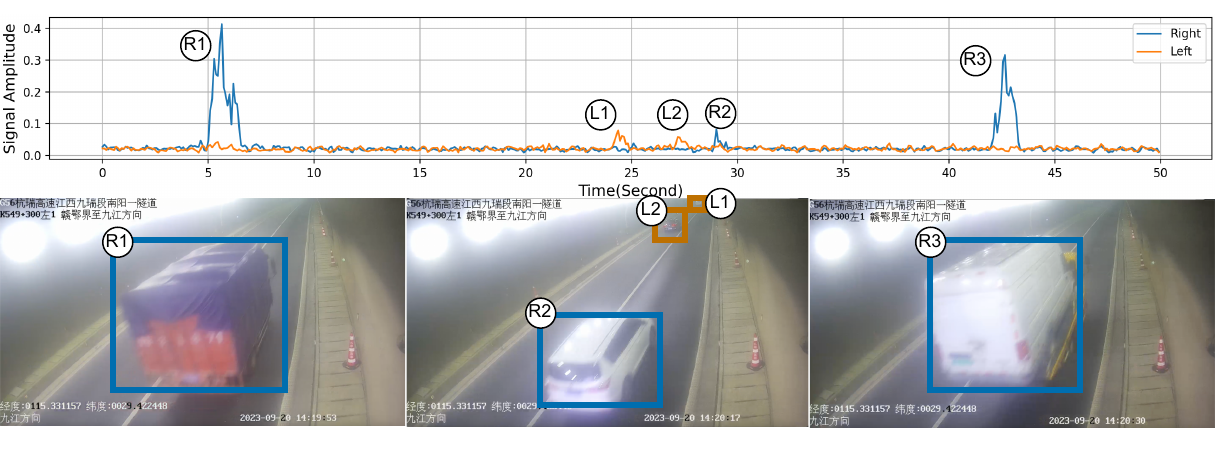}
    \caption{Traffic in a single two-lane tunnel over 50 seconds. The blue and yellow lines show how the amplitude of the fiber changes as vehicles pass by on the right and left, respectively. During this period, a total of five vehicles passed by: three on the right side of the tunnel (R1,R2 and R3) and two on the left side (L1 and L2).}
    \label{F-B_amplitude}
\end{figure*}

Experiments were performed on road-perpendicular offsets $d_y$ to the fiber and vehicle loads $F$, while keeping other parameters constant. As shown in Fig. \ref{F-B}, the simulated impulse responses become stronger as the distance from the laying position decreases and as vehicle load increases. Furthermore, we find these conclusions consistent with actual observations. In our self-established one-way, two-lane tunnel scenarios, we recorded vehicle amplitudes as they passed through. As illustrated in Fig. \ref{F-B_amplitude}, fibers on the left lane only recorded small vehicles L1 and L2, while fibers on right lane recorded R1, R2 and R3 vehicles. This phenomenon demonstrates that fibers primarily record vehicles on each side, verifying that signals decrease with the increase of vertical distance from the fibers. Additionally, the amplitudes of larger vehicles like R1 and R3 are stronger than those of the smaller vehicles (L1, L2 and L3), which confirms that signals increase with the increase of vehicle loads.

\begin{figure}[t!]
    \centering
    \includegraphics[width=3in]{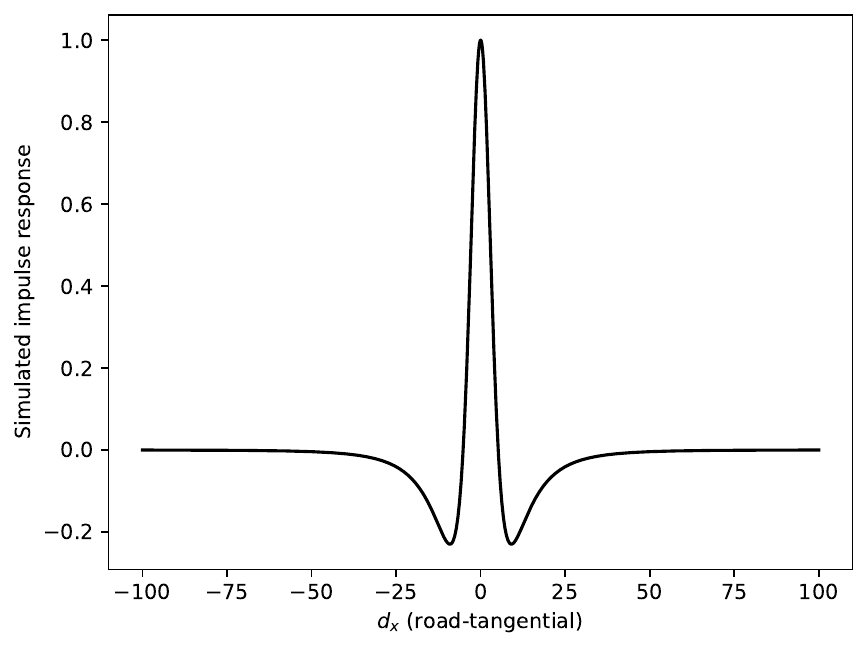}
\caption{Normalized simulated impulse response using practical settings.}
\label{impulse_response}
\end{figure}

According to the parameter settings of the optical fiber acquisition equipment in the real scene of this study, the DAS channel spacing was set equal to the gauge length of 0.8 m and the laying depth is about 5 cm to 10 cm. The one-dimensional normalized simulated impulse response is calculated according to \ref{Simulated DAS response}, as shown in Fig. \ref{impulse_response}, which also serves as the convolution kernel used in Eq. (\ref{lasso_fre}). 

\section{Network Architecture}
\label{architecture}

Due to the insufficiency of annotated data and time-consuming labeling in the supervised denoising algorithms\cite{zhao2023comparison}, we consider adopting a self-supervised learning algorithm. This new learning paradigm is trained from signals generated by data itself, which reduces the dependence on manual annotation and has stronger generalization ability \cite{gui2024survey}. 

\subsection{Architecture}

The proposed hybrid deep-learning network (HDLNet) is consist of a denoising autoencoder (DAE) and a long short-term memory (LSTM). The architecture of this algorithm is an effective U-Net model.

\begin{figure}[!t]
  \centering
  \includegraphics[width=5in]{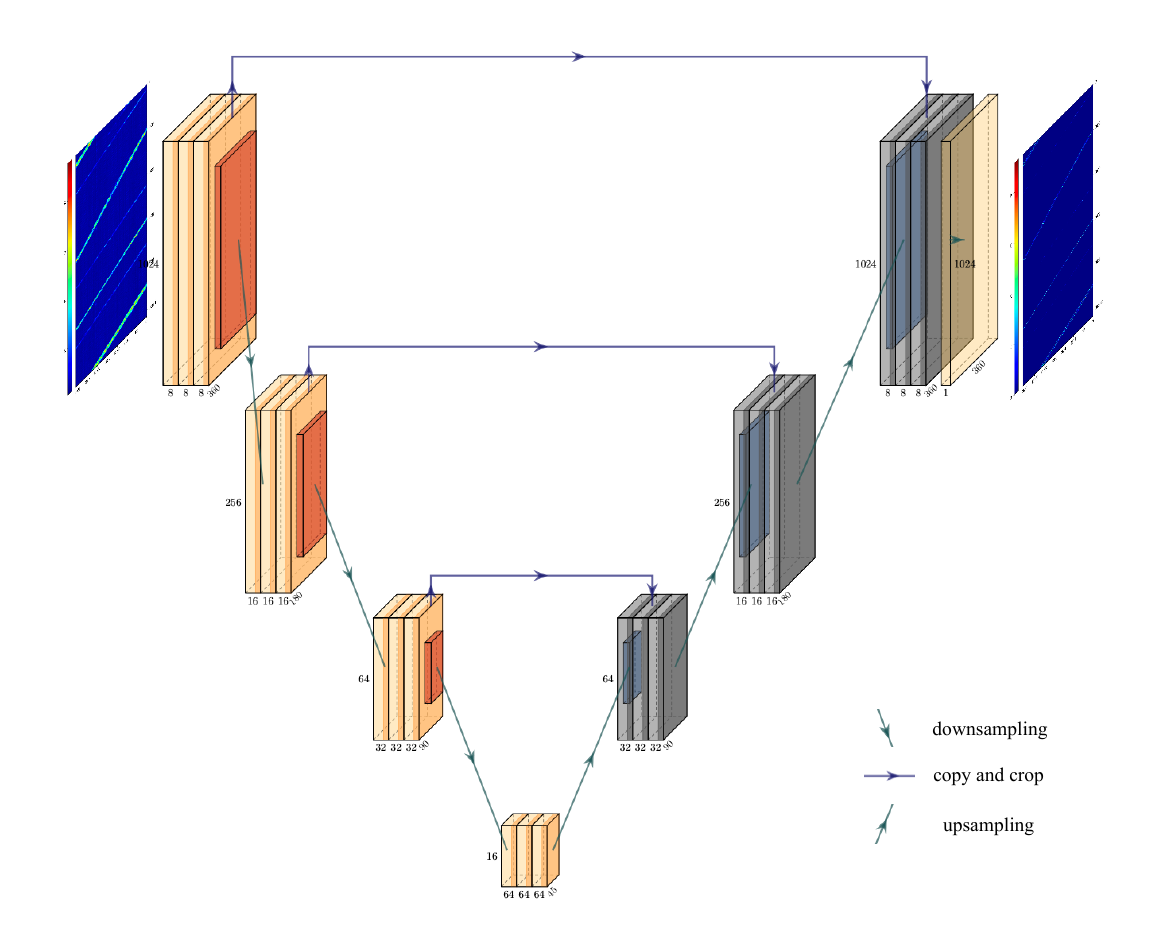}
  \caption{U-Net architecture of the proposed HDLNet. Each box corresponds to a multi-channel feature map. The number of channels is indicated on the bottom of the boxes. The x-y-size is provided the length and width of each box.}
  \label{network}
\end{figure}

\subsubsection{U-Net}

U-Net adopts the U-shaped structure of encoding (downsampling) and decoding (upsampling), which can maintain the same size of input and output \cite{ronneberger2015u}. Besides, this lightweight pixel-level U-Net is particularly suitable for small samples sizes, unbalanced data, and tasks where detailed information needs to be preserved. Due to its huge potential in image denoising, U-Net is used to act as an autoencoder in our DAE model. 

The model inputs are $N_b = 128$ batchs of optical fiber recordings $(y_i \in M_{N_d \times N_t}(\mathbb{R}))_{1 \leq i \leq N_b}$, where $N_d = 360$ is distance samples (288 meters) and $N_t = 1024$ time samples (93.09 seconds). The model outputs $x_i$ is of the same size as $y_i$. The overall U-Net architecture utilized in this paper is shown in Fig. \ref{network}. The network structure is similar to U-shape, whose left and right side are encoding and decoding processes respectively. Encoding layer consists of the repeated application of three $3 \times 5$ convolutions, each followed by a $2 \times 4$ max pooling operation for downsampling. The number of feature channels is doubled after each downsampling operation, starting at 8. Similarly, decoding layer is composed of three repeated $3 \times 5$ convolutions, each ahead of a $2 \times 4$ convolution for upsampling. The number of feature channels is halved after each upsampling operation, starting at 64. Besides, there is a concatenation with the corresponding cropped feature map from decoding layer. Finally, a single $3 \times 5$ convolution with a rectified linear unit (ReLU) activation is used to map each 8-component feature vector to our desired shape.

\subsubsection{LSTM}

Considering the advantages of LSTM in capturing long-term dependencies in time series signals \cite {hochreiter1997long}, it is combined with DAE to improve our HDLNet performance. Following DAE, the input for LSTM network is the output of DAE. Specifically, LSTM first converts the DAE output into a tensor of 360 DAS channels by 1024 samples to ensure the requirenments of the model input. Then process this reshaped data using an LSTM network with the number of units set to 128, configured to return outputs for all time steps in order to maintain the same number of time steps as the input. To further process the output of the LSTM, we apply a fully connected layer containing 1024 neurons to each time step, enabling the network to learn complex features in the sequence data. After processing through the full connected layer, the output is reshaped to match the shape of the original data, thereby providing critical support for the overall performance of our HDLNet.

\subsection{Cost Function}

In LASSO problem Eq. (\ref{lasso}), the model output $x$ cannot be interpreted as an exact reconstruction of the input $y$ , but rather as the parameters (typically the mean) of the conditional probability distribution $P(x|y)$ \cite{vincent2010stacked}. Here, $x|y \sim N(y, \sigma^2\mathbf{I})$, where $N(\cdot, \cdot)$ is normal distribution, $\sigma^2$ is the variance and $\mathbf{I}$ is an identity matrix. This means $x_i|y_i \sim N(y_i, \sigma^2)$ for each real-valued $x_i$. Thus, the conditional probability can be expressed as:
\begin{equation}
    P(x_i|y_i) = \frac{1}{\sqrt{2\pi}\sigma}e^{-\frac{x_ik -y_i}{2\sigma^2}}.
    \label{pro}
  \end{equation}
% This leads to an associated reconstruction error to be optimized:
% \begin{equation}
%   \mathcal{L}(x, y) \propto -{\rm log}P(x｜y).
% \end{equation}

The likelihood function $\mathcal{L}(x, y)$ is the product of each probability density function $P(x_i|y_i)$ for $1 \leq i \leq N_b$:
\begin{equation}
    \mathcal{L}(x, y) = \prod\limits_{i=1}^{N_b}P(x_i|y_i).
\label{probability}
\end{equation}

To facilitate derivation, the likelihood function $\mathcal{L}(x, y)$ is transformed into a logarithmic function $l(x, y)$:

\begin{equation}
    l(x, y) = - {\rm{log}} \mathcal{L}(x, y) \propto \sum\limits_{i=1}^{N_b}(x_ik-y_i)^2
\end{equation}

Furthermore, replace $x_ik$ with $X_i * K$ in the frequency domain and supplemented the regularization term $\lambda \Vert x \Vert _1$ in Eq. (\ref{lasso_fre}), the loss function is expressed as:
\begin{equation}
    l(X, Y) = \frac{1}{N_b} \sum\limits_{i=1}^{N_b}(\Vert X_i * K - Y_i \Vert _2^2 + \lambda \Vert X_i \Vert _1)
\end{equation}

During the training procedure, the parameters are updated to minimize the loss function.

% Due to the maximization of the likelihood function is equivalent to minimizing the mean square error function, the optimal parameter estimation $\hat{\theta}_{DAE}$ of DAE and $\hat{\theta}_{RNN}$ of RNN networks can be given by the following optimization function:

\subsection{Trajectory Extraction}

\begin{figure}[t!]
    \centering
    \includegraphics[width=3in]{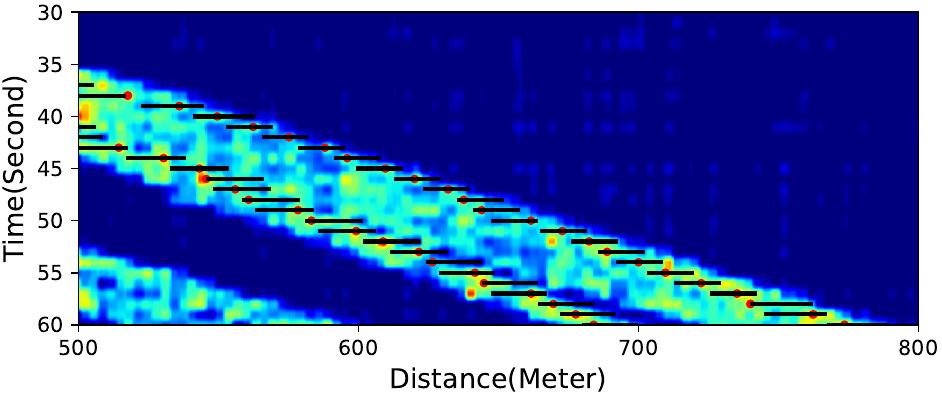}
\caption{An illustration of the trajectory extraction algorithm. The black line segments represent the confidence intervals for distance measurements, while the red dots signify the key points of the vehicle trajectories.}
\label{tunnel}
\end{figure}

In order to further verify the effectiveness of our denoising algorithm for subsequent feature extraction, we utilized our methodology proposed before in \cite{WANG2025104228} for detecting and tracking vehicle trajectories. The flow of the algorithm is illustrated in Algorithm \ref{algorithm_1}. This algorithm can track complete vehicle trajectories and calculate their instantaneous speeds. We evaluated the algorithm using real signals collected from highway tunnel scenarios, as shown in Fig. \ref{tunnel}. The highlighted signals represent the trajectories of vehicles, while the dark blue signals indicate background noise. The first column of red dots marks the collected position information at the moment of vehicle entry. The black lines are combined with the speed limit information of the tunnel section to provide a certain degree of confidence in the position estimates. The subsequent red dots represent the position information of the vehicle recorded every second. Notably, even when two vehicles in close proximity, the algorithm can accurately extract real-time position information for each vehicle. This capability enables the calculation of both average velocity and instantaneous velocity.

% \subsubsection{Vehicle Position Detection}
% After denoising optical fiber signals, we extract the peaks from the first column signals using the peak location search method. This method identifies points where the first derivative is zero and the second derivative is negative. By applying a preset threshold, we calculate the local maximum points and retain only those peak values that exceed this threshold as the vehicles' entry times.

% \subsubsection{Vehicle Trajectory Extraction}
% Based on Newton's first law of motion, a vehicle can be regarded as traveling at a constant velocity over a short period of time. In this paper, the time interval of signals is treated as the minimum unit, and we assume that the velocity of the vehicle remains constant within each minimum unit. After determining the starting position, the peak value in the minimum unit can be obtained line by line in the following time and recorded as the next position point, so that the full tracking of the vehicle can be achieved.

\begin{algorithm}[htbp]
    \caption{Line-by-Line Tracking Algorithm.}
    \label{algorithm_1}
    \LinesNumbered %要求显示行号
    \KwIn{$D = (d_{ij})$: preprocessed optical fiber signal matrix;\ $i = 1, ..., m$, $j = 1, ..., n$, $d_{ij} \in \mathbb{R^{+}}$: selected time and distance.}
  
    \KwOut{$\Sigma^{(S)} = \{(\sigma^{(s)})_{1 \leq s \leq S}\}$: vehicle trajectory point sets.}

    $\{k_s \in \mathbb{N^{+}}, s = 1, ...,S\}$ $\gets$ FindPeak($D_{1: m,1}$), where $S$ is vehicle counts and $k_s$ is the appearance time of each vehicle\;
    $l_{k_s + \alpha}\in \mathbb{N^{+}}$ is distance position point corresponding to time position point $k_s + \alpha$ in $D = (d_{ij})$\;
    $(k_s + \alpha, l_{k_s + \alpha})$ is a pair of time and distance position points of each extracted vehicle trajectory\;

    $\Sigma^{(S)} = \{(\sigma^{(s)})_{1 \leq s \leq S}: \sigma^{(s)} = (k_s + \alpha, l_{k_s + \alpha}), \alpha \in \mathbb{N}$\}\;
    
    Select the initial velocity interval $v_{\min}^{(1)}$ to $v_{\max}^{(1)}$\;
    Calculate the initial distance interval  $x_{\min}^{(1)}$ to $x_{\max}^{(1)}$\;

    \ForEach{$s = 1$ to $S$; $\sigma^{(s)} \in \Sigma^{(S)}$}{
      Add the first pair $(k_s, l_{k_s})$ to vehicle trajectory point sets $\sigma^{(s)}$, where $l_{k_s} = 1$\;
      $l_{k_s + 1}$ $\gets$ ColIndex(Max ($D_{k_s + 1, l_{k_s}+ x_{\min}^{(1)}: l_{k_s} + x_{\max}^{(1)}}$))\;
       \eIf{$k_s + 1 < m$ $\rm and$ $l_{k_s + 1} < n$}{
          Add pair $(k_s + 1, l_{k_s + 1})$ to each vehicle trajectory point set $\sigma^{(s)}$\;
       }{
          break;
       }
      } 

    \ForEach{$s = 1$ to $S$; $\sigma^{(s)} \in \Sigma^{(S)}$}{
      \For{$\alpha \in \mathbb{N^{+}}$}{
        $y^{(s)}$ = PolynomialFitting($\sigma^{(s)}$)\;
        $v^{(s)}$ = Slope($y^{(s)}$)\;
        Select the right level of confidence $\rm cof$\;
        Calculate velocity interval $(1 + \rm cof)$$v^{(s)}$ to $(1 + \rm cof)$$v^{(s)}$\;
        Calculate distance interval $x_{\min}^{(s)}$ to $x_{\max}^{(s)}$\;
        $l_{k_s + \alpha}$ $\gets$ ColIndex(Max ($D_{k_s + \alpha, l_{k_s + \alpha - 1}+ x_{\min}^{(s)}: l_{k_s + \alpha - 1} + x_{\max}^{(s)}}$))\;
       \eIf{$k_s + \alpha < m$ $\rm and$ $l_{k_s + \alpha} < n$}{
          Add pair $(k_s + \alpha, l_{k_s + \alpha})$ to each vehicle trajectory point set $\sigma^{(s)}$\;
       }{
          break;
       }
  
      } 

      }
    \textbf{Return:} Vehicle trajectory point sets $\Sigma^{(S)}$.
\end{algorithm}

\section{Experiments}
\label{experiments}

\subsection{Datasets}

\begin{figure}[t!]
    \centering
    \includegraphics[width=5in]{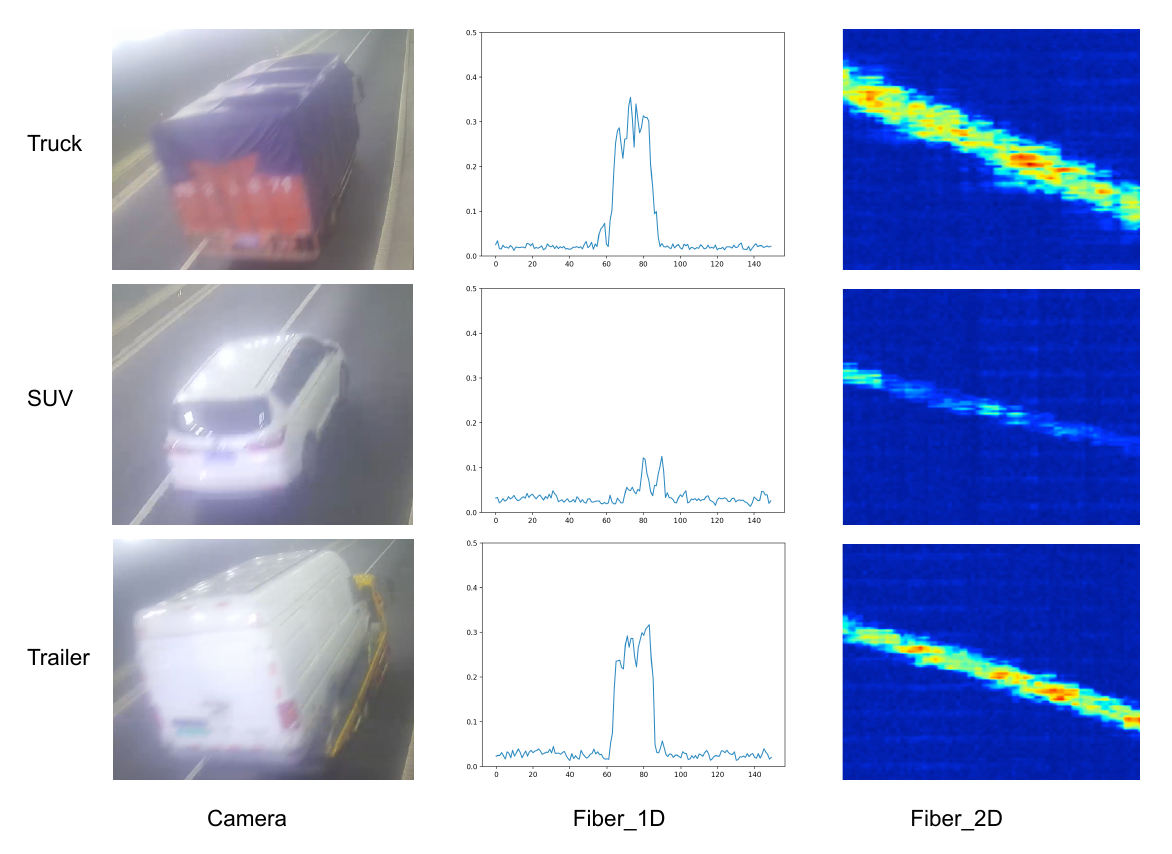}
\caption{Examples in Jiurui tunnel dataset with three types like truck, SUV and trailer. Cameras and their corresponding one-dimensional fiber signal amplitudes and two-dimensional waterfall diagrams are shown for display.}
\label{dataset}
\end{figure}

To our best knowledge, there is no publicly available optical fiber dataset focusing specifically on highway tunnel scenes. To bridge this gap and provide a robust basis for evaluating our proposed algorithms, experiments were conducted in real highway tunnel scenarios. The Jiurui tunnel dataset, located in Jiujiang City, Jiangxi Province, China, represents a real-world scenario featuring a single carriageway with two lanes. In our DAS system, the time interval for collecting data samples at each acquisition point is set to 1/11 seconds, equating to 11 samples collected per second. A total of 99,352 fiber signals were recorded in the Jiurui tunnel dataset, corresponding to approximately 2.5 hours of data collection. The DAS acquisition configuration are shown in Table \ref{configuration}. Detailed descriptions of the samples are presented in Fig. \ref{dataset}.

\begin{table}[!t]
    \caption{DAS recording parameters.}
    \label{configuration}
    \centering
    \renewcommand\arraystretch{1.2}
    \begin{tabular}{cc}
    \hline
    Parameter & Jiurui tunnel dataset\\
    \hline
        Pulse repetition frequency & 4000Hz \\
        Gauge length & 5000m \\
        Sample spacing & 80cm \\
        ADA sampling frenquency & 4Gbps \\
        \hline
    \end{tabular}
\end{table}

\subsection{Implementation details}

The proposed network is implemented in a Python 3.8 environment with TensorFlow 2.10.0, running on a Linux operating system, utilizing a single Nvidia GeForce RTX 3060 GPU. During the training procedure, we employed the Adam optimizer \cite{kingma2014adam} with a learning rate of 0.0005 and a batch size of 128. The total number of parameters in the proposed hybrid model is 9,747,393, and we trained the model for 1,000 epochs using simulated spatial impulse responses, as depicted in Fig. \ref{impulse_response}. These parameters are aligned with our real settings from the Jiurui Tunnel experiments. 

From Jiurui tunnel dataset dataset, we partitioned 30 minutes for evaluation, while the remaining 2 hours were designated for training. The training set and test set were divided in a ratio of 80\% to 20\%, enabling to generate synthetic data for both training and validation purposes. Before the training and testing procedures, the vibration intensity of the signals was normalized to the range of [0,1] to ensure consistency and enhance the performance of our algorithms.

\subsection{Metrics}

To further quantify the reconstruction results from the original to the target, the hybrid network is evaluated using mean squared error (MSE) and peak signal-to-noise ratio (PSNR), defined by the following formulas:
\begin{equation}
    {\rm MSE} = \frac{1}{n}\sum\limits_{i = 1}^{n}(y_i - \hat{y}_i)^2
\end{equation}
\begin{equation}
    {\rm RSNR} = 10{\rm log}\frac{v^2}{\rm MSE}
\end{equation}
where $y_i$ represents the original signals, $\hat{y}_i$ denotes the reconstructed signals, $n$ is the total number of sampling points, and $v$ is the maximum value of the image pixels, which is 255 for an 8-bit image (i.e., $2^8 - 1$). Generally, lower MSE values and higher PSNR values indicate better reconstruction performance.

However, the aforementioned evaluation indices do not adequately measure the similarity of two images as perceived by the human eye. To address this, the  structural similarity index (SSIM) has been introduced, based on the premise that the human visual system can extract structured information from images, making it more aligned with human visual perception compared to traditional methods \cite{wang2004image}. SSIM is a full-reference evaluation index that considers brightness, contrast and structure of images. It is defined as follows:
\begin{equation}
    {\rm SSIM} = [l(y, \hat{y})]^\alpha[c(y, \hat{y})]^\beta[s(y, \hat{y})]^\gamma
\end{equation}
where $l(y, \hat{y}), c(y, \hat{y})$, and $s(y, \hat{y})$ represent luminance, contrast, and structure, respectively. These components are calculated as: 
$$
l(y, \hat{y}) = \frac{2\mu_{y}\mu_{\hat{y}} + c_1}{\mu_y^2 + \mu_{\hat{y}}^2 + c_1}
$$

$$
c(y, \hat{y}) = \frac{2\sigma_{y\hat{y}} + c_2}{\sigma_{y}^2 + \sigma_{\hat{y}}^2 + c_2}
$$

$$
s(y, \hat{y}) = \frac{\sigma_{y\hat{y}} + c_3}{\sigma_{y} + \sigma_{\hat{y}} + c_3}
$$
where $\mu_{(\cdot)}, \sigma_{(\cdot)}$, and $\sigma_{(\cdot, \cdot)}$ refer to mean, standard deviation, and covariance, respectively, while $c_{(\cdot)}$ are constants. The SSIM value ranges from -1 to 1, where values closer to 1 indicate a superior denoising effect.

\subsection{Analysis}

Due to the unique characteristics of optical fiber signals, they can be represented as either one-dimensional signal diagrams or two-dimensional waterfall diagrams. Additionally, based on the denoised images, we examine the performance of the proposed line-by-line matching algorithm for vehicle identification and tracking. This comprehensive analysis aims to fully demonstrate the processing results achieved by our algorithm.

\subsubsection{One-Dimensional Signal Analysis}

In the analysis of one-dimensional signal amplitudes, we present 5 minutes of optical fiber signals collected from a fixed location in the test set. Additionally, we conducted comparative experiments between the proposed algorithm and Spatial-DAE \cite{yuan2023spatial} using the same datasets and model parameters. 

\begin{figure}[t!]
    \centering
    \subfigure[]{
        \centering
        \includegraphics[width=3.5in]{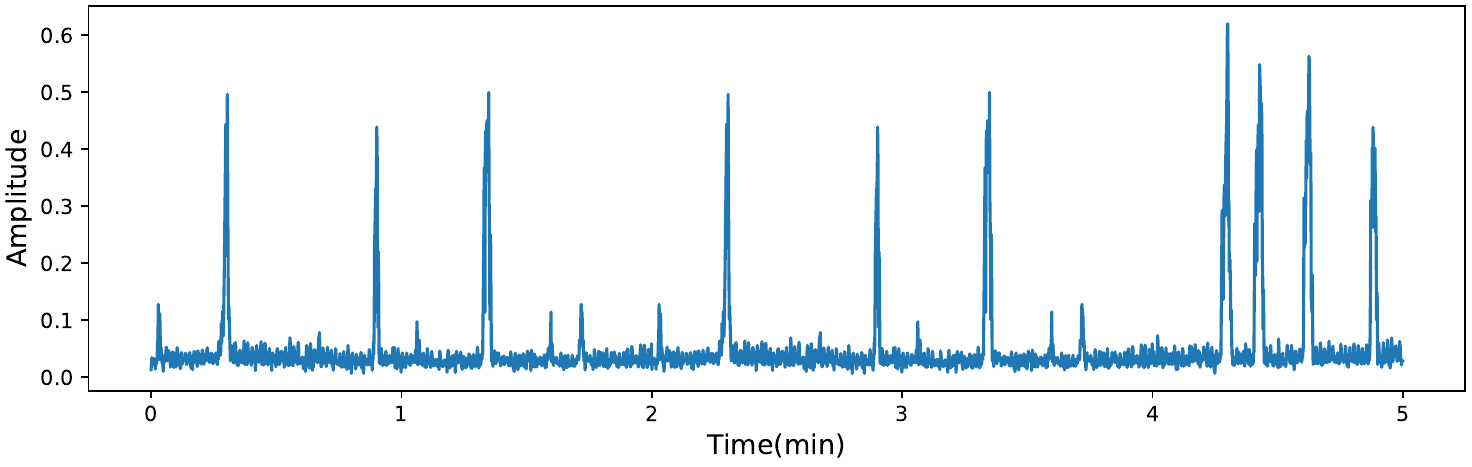}
    }\hfill

    \subfigure[]{
        \centering
        \includegraphics[width=3.5in]{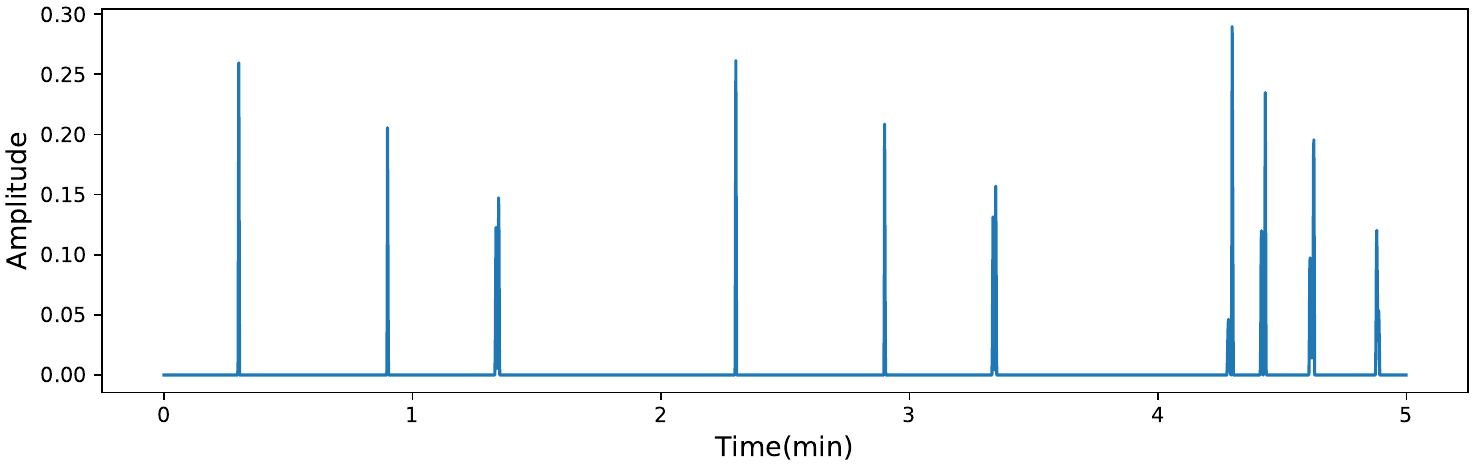}
    }\hfill

    \subfigure[]{
        \centering
        \includegraphics[width=3.5in]{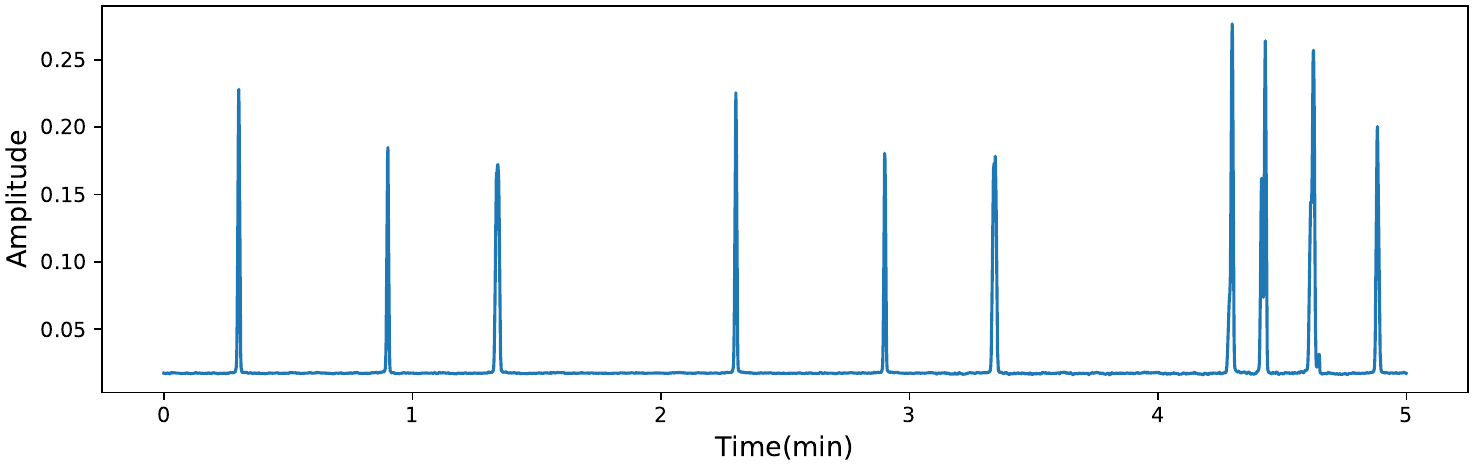}
    }\hfill
\caption{1-D optical fiber amplitudes from the Jiurui Tunnel dataset over a 5-minute duration. (a) Raw signals, (b) denoising results using Spatial-DAE, and (c) denoising results using the proposed algorithm.}
\label{1d}
\end{figure}

As shown in the visualization results in Fig. \ref{1d}, both Spatial-DAE and our proposed algorithm effectively remove small peaks of non-vehicle amplitudes present in the background noise, successfully retaining the vehicle amplitudes with accurate coordinates. Moreover, our algorithm has demonstrated superior capability in characterizing the wave width information of vehicles. Furthermore, the amplitude intensity of the vehicles identified by our algorithm generally aligns with the intensity of the original signals, whereas Spatial-DAE tends to produce intermodulation effects between strong and weak amplitudes.

\subsubsection{Two-dimensional signal analysis}

\begin{figure*}[t!]
    \centering
    \subfigure[]{
    \includegraphics[width=1.6in]{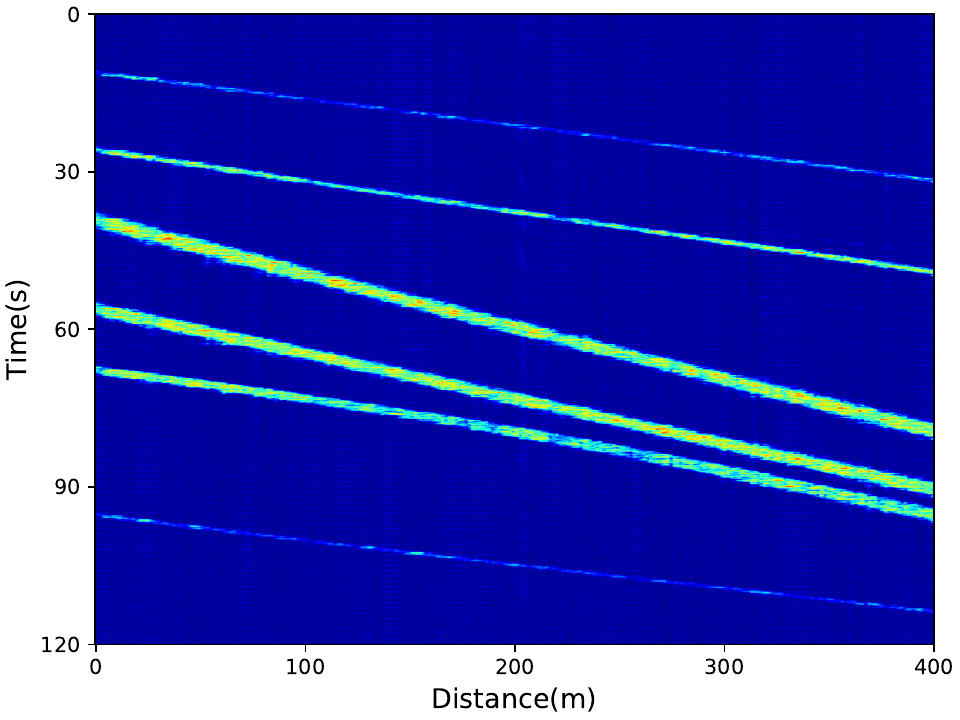}
    }
    \subfigure[]{
    \includegraphics[width=1.6in]{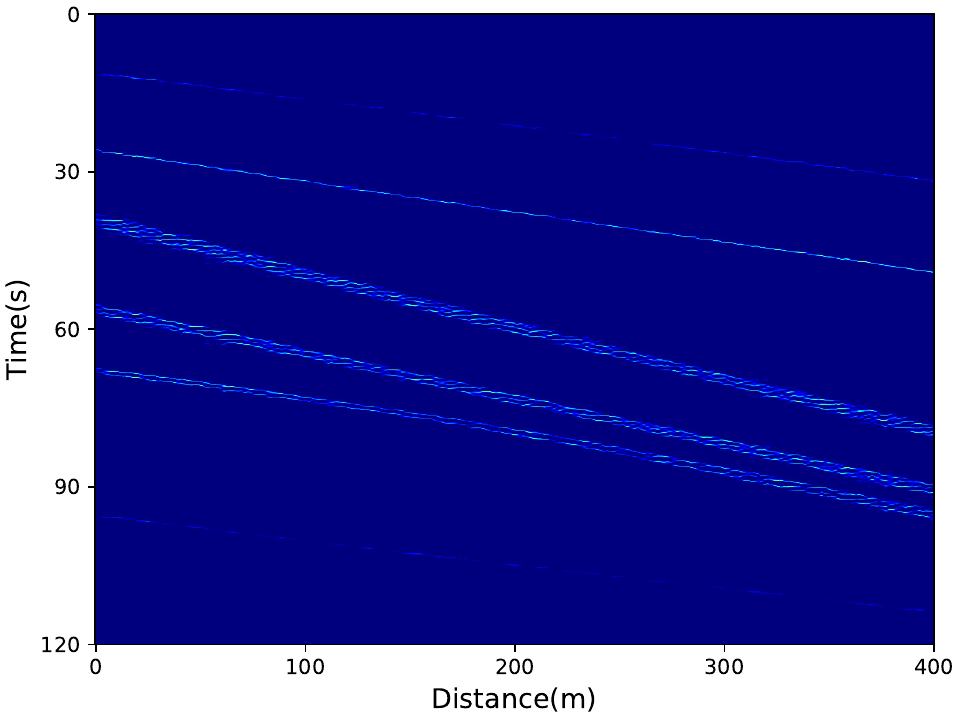}
    }
    \subfigure[]{
    \includegraphics[width=1.6in]{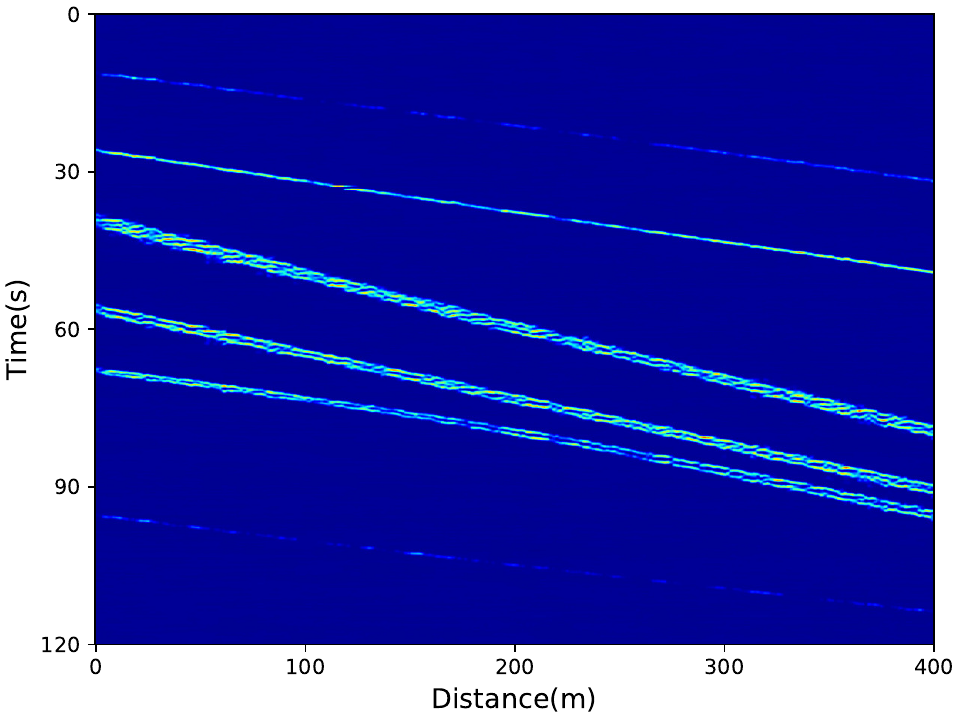}
    }

    \subfigure[]{
        \includegraphics[width=1.6in]{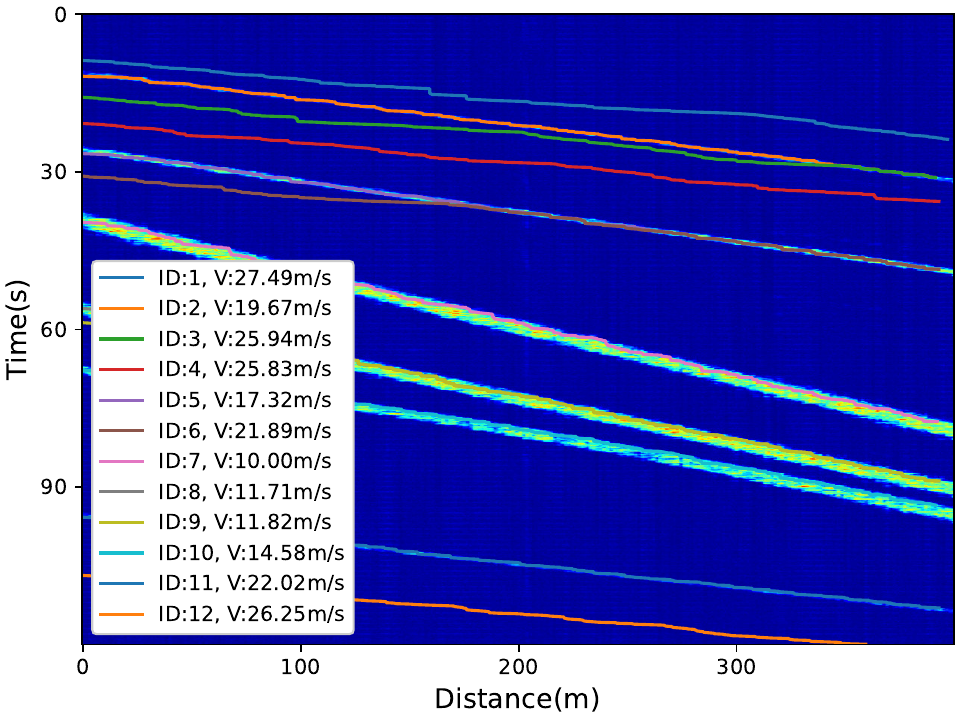}
        }
    \subfigure[]{
        \includegraphics[width=1.6in]{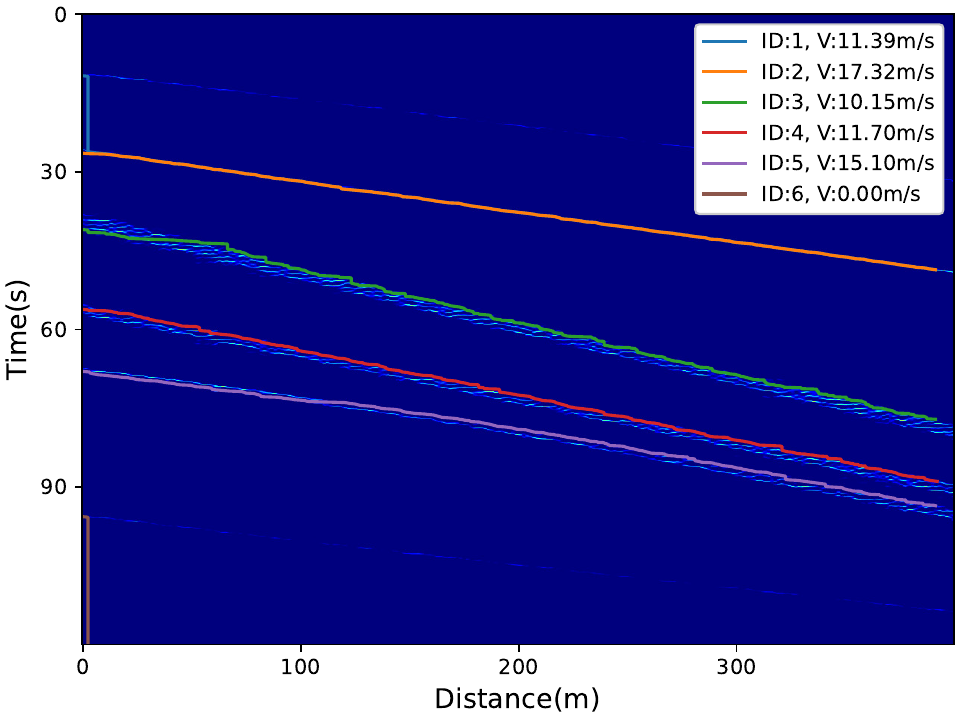}
        }
    \subfigure[]{
        \includegraphics[width=1.6in]{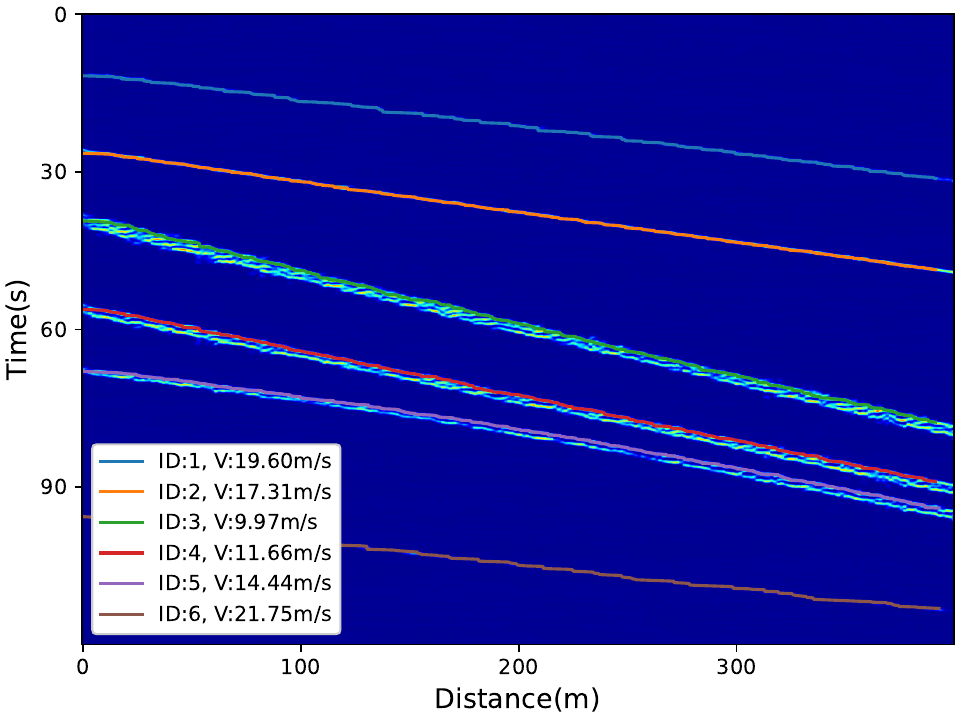}
        }
\caption{2-D optical fiber waterfall diagrams from the Jiurui Tunnel dataset over a 400-meter 2-minute section. (a), (b) and (c) show raw signals, denoising results using Spatial-DAE and our HDLNet respectively. (d), (e) and (f) display corresponding vehicle trajectory extraction results.}
\label{2d}
\end{figure*}

In the analysis of 2-D signal amplitudes, we presented optical fiber signals over a duration of 2 minutes and a distance of 400 meters from the test set. As shown in the visualization results in Fig. \ref{2d}(a)(b)(c), the proposed algorithm effectively removes background noise while retaining clear driving trajectories. In contrast, Spatial-DAE struggles to recover the trajectories of small vehicles and fails to adequately represent the trajectories of larger vehicles. Here, we also utilized the line-by-line matching algorithm with the same parameters to verify the effectiveness of the denoising algorithms for subsequent feature extraction. As shown in Fig. \ref{2d}(d)(e)(f), raw diagrams without produce too much noise, and denoised diagrams by Spatial-DAE are difficult to display small car information. 

Building on the visualized results displayed in the waterfall diagrams, we conducted a quantitative analysis. As detailed in Table \ref{quantitative}, the proposed denoising algorithm achieves smaller MSE values and larger PSNR and SSIM values, indicating superior performance compared to the Spatial-DAE method.

\begin{table}[!t]
    \caption{Quantitative indexes of different algorithms in tunnel scenarios.}
    \renewcommand\arraystretch{1.2}
    \centering
    \begin{tabular}{cccc}
    \hline
    Method & MSE & PSNR & SSIM \\
    \hline
    Spatial-DAE & 0.1148 & 18.7970 & 0.7524 \\
    Proposed Algorithm &  \textbf{0.0829} & \textbf{21.6280} & \textbf{0.8201} \\
    \hline
    \end{tabular}
    \label{quantitative}
\end{table}

\subsubsection{Trajectory Extraction Analysis}

\begin{figure*}[t!]
    \centering
    \subfigure[]{
        \centering
        \includegraphics[width=5in]{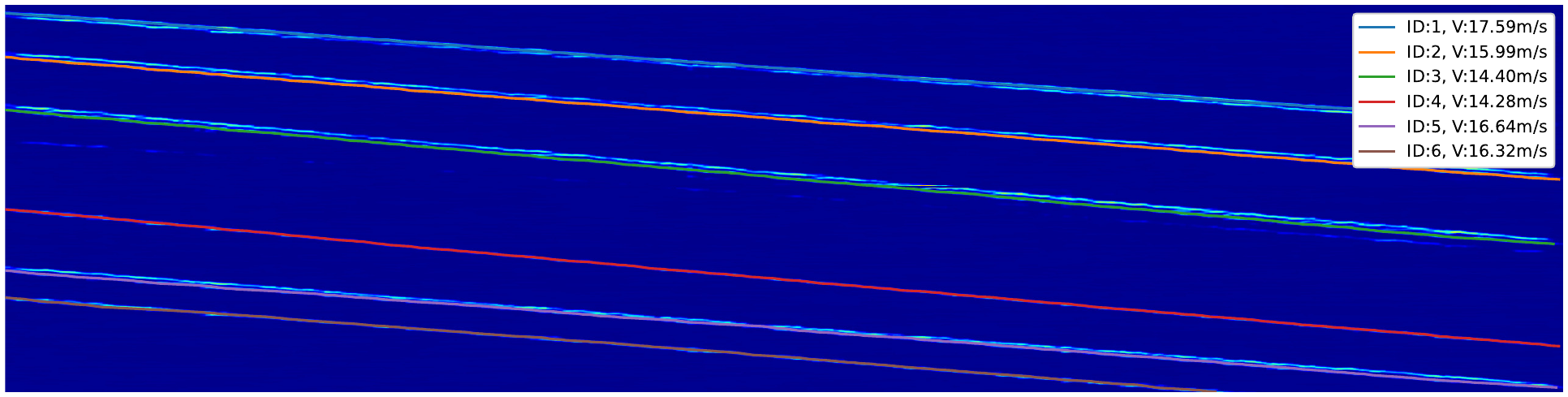}
    }\hfill
    \subfigure[]{
        \centering
        \includegraphics[width=5 in]{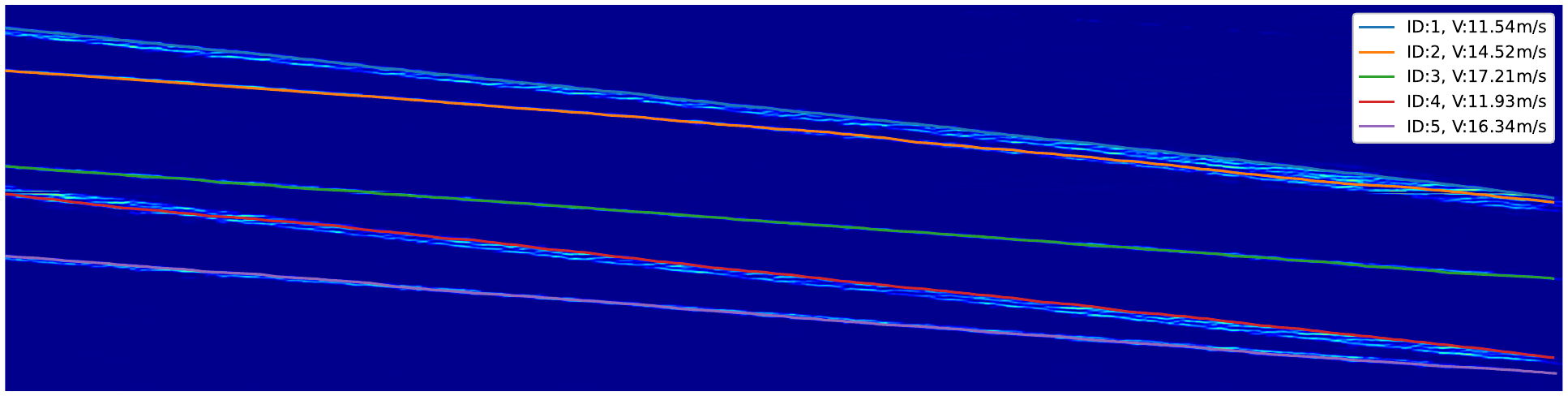}
    }\hfill
    \caption{Trajectory and velocity extraction results using the line-by-line algorithm based on proposed HDLNet denoised diagrams. (a) Vehicles traveling at roughly constant speeds, (b) Vehicles traveling at variable speeds.}
    \label{trajectory}
\end{figure*}

In addition to the qualitative and quantitative analyses presented above, we also carried out feature extraction on the denoised signals to verify the effectiveness of the denoising algorithms. The previously proposed line-by-line matching algorithm in \cite{WANG2025104228} is applied for vehicle detection and tracking. Different traffic conditions are selected for display. As illustrated in Fig. \ref{trajectory}(a), six vehicles are traveling at roughly constant speeds, while the vehicles depicted in Fig. \ref{trajectory}(b) exhibit slightly different and variable speeds. The line-by-line matching algorithm can effectively recognize and track vehicles without encountering ID-switching issues. Additionally, it enables the calculation of average speeds, which can be utilized for subsequent traffic statistics. It can be seen that our denoising algorithm is beneficial to the subsequent algorithm processing.

\section{Conclusion}
\label{conclusion}
In this paper, we developed a comprehensive set of algorithms for vehicle detection using optical-fiber based distributed acoustic sensing (DAS) systems. To facilitate practical applications on real highways, we constructed and presented a dataset specifically for tunnel scenarios. Additionally, we validated the effectiveness of the Flamant-Boussinesq model in characterizing fiber signals, demonstrating that signal amplitude increases with decreasing road-perpendicular offsets and increasing vehicle loads.

For data preprocessing, we proposed a self-supervised deep learning network named HDLNet. This algorithm integrates the denoising capabilities of a denoised autoencoder (DAE) with the sequential processing capabilities of a long short-term memory (LSTM), achieving satisfactory reconstruction results. Through visualization and quantitative analysis, our algorithm outperformed the previously proposed Spatial-DAE model across various metrics, highlighting the advantages of incorporating LSTM. Besides, our algorithms have the potential for application in spatio-temporal signal processing across other fields.

Furthermore, we also verify the effectiveness of the denoising algorithm for subsequent vehicle extraction algorithms. Based on the denoised optical fiber waterfall diagrams, we utilized a line-by-line matching algorithm for vehicle detection and real-time tracking. The experiment verifies that the denoised diagrams by HDLNet can achieve better feature extraction effect than original diagrams and the denoised diagrams by Spatial-DAE.

On the whole, we demonstrated the feasibility of optical fiber-based DAS systems within the context of intelligent transportation systems (ITS) for highway tunnel traffic monitoring. Additionally, the signal processing scheme proposed in this paper may facilitate the implementation of related technologies in the industry.

\section*{Declaration of Competing Interest}

The authors declare that they have no known competing financial interests or personal relationships that could have appeared to influence the work reported in this paper.

\printcredits

\section*{Acknowledgments}

This research was funded by Tianjin Yunhong Technology Development (Grant number: 2021020531).

% \bibliographystyle{cas-model2-names}
\bibliographystyle{elsarticle-num-names.bst}

% \bibliographystyle{unsrt}

\bibliography{reference}

\end{document}